# W4IPS: A Web-based Interactive Power System Simulation Environment For Power System Security Analysis

Zeyu Mao
Texas A&M University
zeyumao2@tamu.edu

Hao Huang
Texas A&M University
hao_huang@tamu.edu

Katherine Davis
Texas A&M University
katedavis@tamu.edu

## Abstract

*Modern power systems are increasingly evolving cyber-physical systems (CPS) that feature close interaction between information and communication technology (ICT), physical and electrical devices, and human factors. The interactivity and security of CPS are the essential building blocks for the reliability, stability and economic operation of power systems. This paper presents a web-based interactive multi-user power system simulation environment and open source toolset (W4IPS) whose main features are a publish/subscribe structure, a real-time data sharing capability, role-based multi-user visualizations, distributed multi-user interactive controls, an easy to use and deploy web interface, and flexible and extensible support for communication protocols. The paper demonstrates the use of W4IPS features as an ideal platform for contingency response training and cyber security analysis, with an emphasis on interactivity and expandability. In particular, we present the use cases and the results of W4IPS in power system operation education and security analysis.*

## 1. Introduction

The emergence of intelligent electronic devices (IEDs) and smart grid technologies has been a catalyst for power grids to evolve over the past decades from a predominately physical infrastructure into a more complicated cyber-physical system (CPS) whose efficiency, reliability and controllability have been greatly improved. The trend is toward more generic interoperable architectures for system-based wide area monitoring, protection and control (WAMPAC), where various applications, including capturing and analyzing dynamic event data, determining real-time system state, designing protection systems for wide-area disturbances, etc., are described in [1]. These applications support improved operational reliability (previously called security) of the power system against physical outages. While these changes should increase power grid resilience, the technologies heavily rely on cyber components and networks. Cyber network compromise can lead to hazards in the power grid that adversely affect the daily lives of citizens, like the Ukraine power grid cyber attack which affected up to 225,000 customers in three different distribution-level service territories [2]. Thus, it is an urgent task to analyze power system security from all aspects.

The importance of situational awareness (SA) for power systems is presented in [3], where Panteli and Kirschen also present the difficulties in quantifying SA, including the size of the grid, the volume of data, and the complexity of operations in modern power grids that prevent sufficient SA to operators. Cyber security is an important concern for SA since the data shown must be available and trustworthy. At the same time, SA is a necessary precursor for making both physical and cyber operating decisions. It is thus important to visualize cyber-physical power systems vividly and reflect key status information in a real-time, easy-to-comprehend manner. Human factors are widely recognized as important, but remain a largely ignored element for power system security analysis. Examples that impact power grid security through SA include the following: *(1) What if the operator, through error or lack of experience, issues a wrong command? (2) What if there is an insider in the control room or control system who intentionally sabotages the system?* As discussed in [4], operator decision making should be considered in the design phase to evaluate potential new applications and visualizations used in power systems operations. Therefore, introducing a scientific mechanism for incorporating human factors into the visualization tools and the associated power system security studies is significant.

In [5], Pienta et al. introduce a spatio-temporal visualization using time-varing geospatial point data with a linear power systems representation to provide users a quick and easy way to understand a power network's behavior over time. Mohapatra and Overbye present dynamic mode decomposition based visualization for large-scale power systems, whose main benefit is fast computation, enabling automated processing and tracking of transient contingency results in an on-line mode [6]. Idehen et al. present the visualization of large-scale electric grid oscillation modes for improved understanding of power system dynamics in [7]. While these methodologies present innovative visualizations for power systems studies and analysis, they do not introduce human factors into the power system analysis. To automatically assess and incorporate human factors, and translate those into improved grid decision-making, a special environment with interactive communication among the power system simulation, visualizations, and users is required.

Ciraci et al. introduce a framework for power system and communication networks co-simulation, FNCS, in [8]. In [9], Huang et al. incorporate the dynamic simulation with FNCS through a decoupled simulation approach to link with existing dynamic simulators. However, the FNCS focuses more on the interaction between transmission and distribution, and the synchronization between communication networks and power system simulation, but not the interactive actions between human operators and power systems. Overbye et al. in [10] presents an interactive power system dynamic simulator for the PMU time frame. This interactive transient stability server enables the agnostic use of third-party computational clients, algorithms, and humans to directly impact the dynamically simulated grid behavior. This innovation also enables the introduction of human factors into power system studies for the period of real-time simulation. Based on this dynamic simulation server, we introduce a web-based interactive power system simulation environment (W4IPS) that can be used with transient and steady-state simulations. Key features of W4IPS include its publish/subscribe architecture, real-time data push and visualizations, interactive controls, multi-user interaction, web interface access, and flexible and extensible communication protocols.

1. The publish/subscribe structure allows instantaneous, push-based data delivery, eliminating the need for data consumers to periodically check or poll for new data and updates. This structure improves the efficiency, performance, reliability and scalability of the underlying communication system.

2. Web interfaces do not require installing dependencies and environments, thus it minimizes the time for deploying the environment for multiple PCs and it can be run in different operation systems.

3. Multi-user interactive control allows multiple users to access the W4IPS and control the devices, which can be used to test different attack and defense strategies in W4IPS.

4. Real-time data visualization and event alert enables users to respond in time under different contingencies.

5. Flexible and extensible communication protocols enable W4IPS to communicate with Internet-of-Thing (IoT) devices in the power grid.

With these features, the paper presents two applications based on W4IPS, including a lab course for teaching undergraduate students about power system operation and a contingency scenario for a Geomagnetic Disturbance (GMD) event in power systems. We also discuss how the W4IPS could be used for cyber-physical security analysis and how the interactivity between human and the grid could be introduced with W4IPS.

The rest of this paper is organized as following. The architecture and methodology of W4IPS is introduced in Section 2. Applications of W4IPS are discussed in Section 3. More discussions of the W4IPS future work is presented in Section 4.

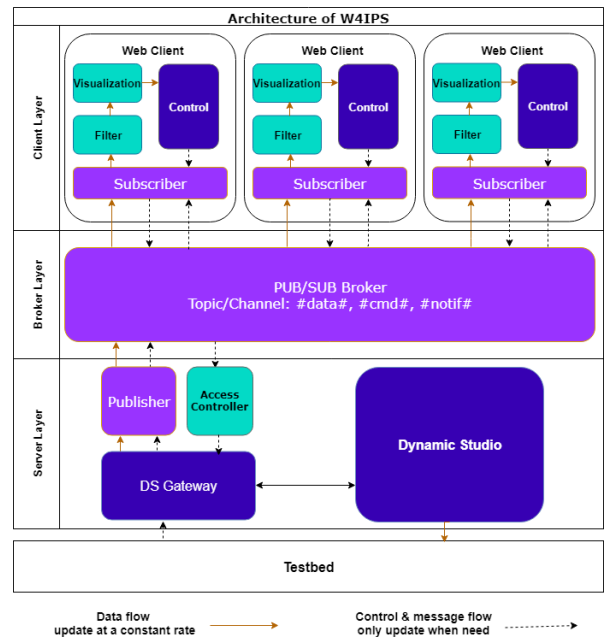

**Figure 1. Architecture of W4IPS:** Client Layer is for users to monitor and control the devices in the simulated grid; Broker layer is to route messages; Server layer serves as a protocol gateway between the PowerWorld DS/c37.118.2 and MQTT protocol

## 2. W4IPS

W4IPS is an open source extensible web-based tool whose design enables distributed multi-user simulation interactions and visualizations. The architecture, data sources, and interfaces of W4IPS are presented below.

## 2.1. Data Source

A significant data source for W4IPS is a power system model with interactive simulation capability. For the transient stability time frame, W4IPS interacts with PowerWorld Dynamics Studio (PW-DS) [10] [11], which is an interactive simulation environment that allows users create events during a simulation. All changes can be stored for later playback of the simulation [10].

PW-DS is commercial software for time-domain, positive sequence power system simulation with Phase Measurement Unit (PMU) time frame or uniform frequency model. The PW-DS can represent the various power system dynamic models, import and export case models in industry standard formats, and efficiently solve large power system cases to provide power system dynamic data [11]. It can be run either in a stand-alone mode for real time visualization or in a server mode for different clients using different communication protocol. For the steady state time frame, W4IPS interacts with PowerWorld Simulator via the SimAuto, which is a callable interface that supports external commands [12].

## 2.2. Communication System

To transmit the significant amount of real-time data among the data source and multiple users, two major factors are taken into serious consideration:

1. Latency. The interaction needs low latency to give users enough time to respond when there is an event happening in the simulation.

2. Bandwidth usage. The bandwidth usage has to be low to leave enough bandwidth for simultaneous data streams.

According to the factors above, Message Queuing Telemetry Transport (MQTT) [13] [14], a lightweight publish/subscribe messaging transport protocol, is selected to serve as the core of the communication system. Three features makes it fit the system requirements:

1. A small transport overhead and protocol exchanges minimized to reduce network traffic.

2. Use of the publish/subscribe message pattern which provides one-to-many message distribution and decoupling of applications.

3. The protocol runs over TCP/IP, or over other network protocols that provide ordered, lossless, bi-directional connections.

4. Same message structure with different topic headers to distinguish the message types. The topic "data" is used for state and variable sharing, the topic "command" is for user actions, and the topic "notif" is for system-wide notification/event broadcasting.

Compared to the commonly used web interaction method, Representational State Transfer (REST), which is built upon HTTP, the MQTT message is 2.5 times smaller on average and the latency is only 1/4 of the level of REST [15]. By tweaking the parameters in the broker and gateway, scenarios considering communication delays and failures could also be achieved. With the support from the MQTT, the real-time multi-user data sharing mechanism can be achieved in the simulation environment.

## 2.3. Architecture

A key aspect of W4IPS is its focus on the capability to have multiple users operating the power grid with the real-time data stream. Under a limited bandwidth, the traffic congestion can be a problem when involving a real-time multi-user scenario, especially when the simulation case grows from a few nodes to thousands of nodes, the transportation of a single data stream can take a significant time and thus the connection consistency is hard to guarantee for multiple users. **Figure 1.** shows that in order to be capable and scalable for this multi-user simulation, instead of using the traditional end-to-end client/server protocol, W4IPS uses a lightweight publish and subscribe messaging exchange protocol for the internal communication, which is used in many real-time applications like online chat/message systems. Each data message in W4IPS has a topic header which decides the routing path of the message.

As shown in **Figure 1.**, according to the usage of the component, W4IPS can be divided into three layers. The client layer consists of multiple web clients, which can be used by operators for power system monitoring and interactive control actions. Normally they are the subscribers of the "data" topic, and only when the operator issues a command, a message with the "command" topic will be published by the web client. Based on the user's role in the scenario, the data that the user has no privilege to view will be filtered. The server layer has a gateway that translates the PowerWorld Dynamics Studio protocol/c37.118.2 into the MQTT protocol and vice versa. When needed, the gateway can be used to connect SimAuto and the MQTT client for the steady state simulation, and is extendable for other protocols. It also has an access controller which determines whether the user has the privilege to execute the attempted control action based on the user's role, e.g. a load shedding command from a user who is responsible for reactive power support will be ignored and tagged as an invalid and suspicious action. There is another layer in the middle, where a broker filters published messages based on topic, and then distribute them to subscribers, transferring real-time data, control commands and notifications between clients and DS.

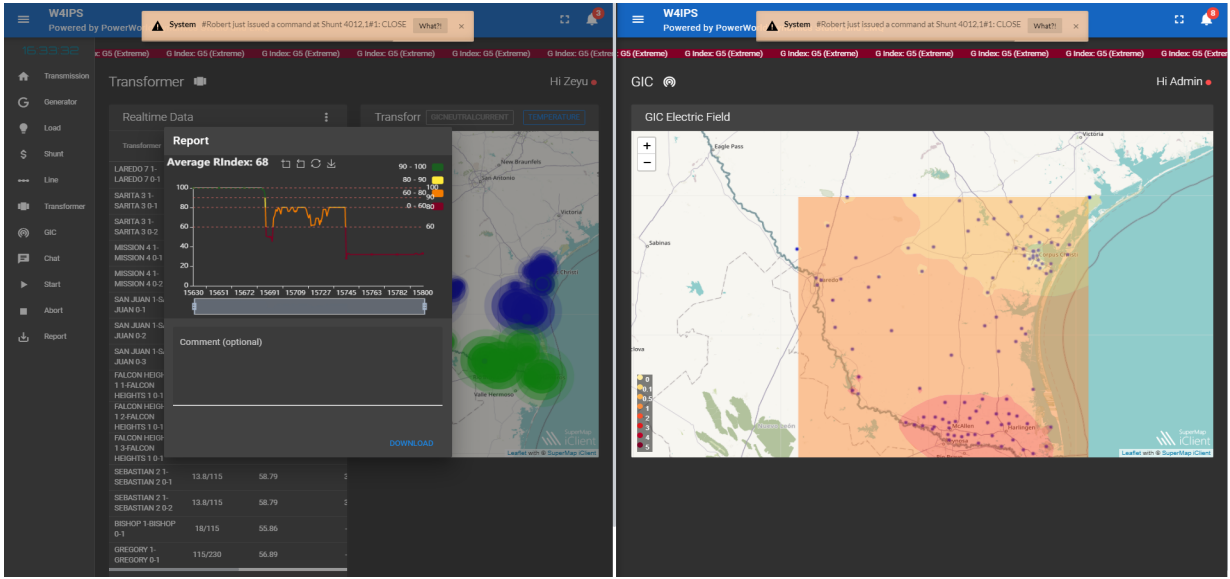

Figure 2. The web interface screenshots. The left one shows an evaluation report popup on top of the transformer table and map. The right one shows a real-time GIC field contour on the map. On the top of both screenshots there are notifications that will show up on each user's screen when someone issues a command.

Compared to the traditional polling method where web clients periodically send requests for data and DS sends out the data based on the query, this publish/subscribe structure largely reduces the number of query, which promotes faster response time, increase the number of concurrent users and reduces the delivery latency that can be particularly problematic in a real-time monitor and control system. As a result, W4IPS has the capability to have 16 users operating a 2000 bus case at the same time. Theoretically, it will have larger capability when working with smaller cases.

One common problem for the multi-user simulation is the interaction convergence. When multiple clients exchange information with the main simulation in the server layers, the values of the variables/states may need to be updated. In the simulation process, this means there is always 1-step mismatch between the present simulation with old variables and the true simulation with true variables. However, in W4IPS, the states are synchronized across all clients by only accepting the data stream from the simulation engine. The simulation engine will not accept the state changes until it solves the next time step based on the changes. For instance, when one client changes the terminal voltage of a generator, the command will be sent to the simulation engine through the communication system. If the command is valid, then the simulation engine will solve the system based on the new state. Then all clients will update both the state and variables according to the latest values.

## 2.4. User Interface

Another feature of the W4IPS is the web-based interface that allows users to use any browser to get access to the system, without installing dependencies, which also minimizes the time cost for deploying a multi-user environment. The interface has customized dashboards for different roles where operators adjust the generation, the status of the shunts in the system, or use a holistic view of the system to make high level decisions for other collaborators. As shown in **Figure 2.**, the interface can be switched by clicking the button in the left panel. Based on the application, the interface may have some additional dashboards:

1. **Security Analysis**: This dashboard includes different power system security analysis, such as power system state estimation [16], cyber security analysis and visualization [17], power system equipment risk assessment [18, 19], etc. This dashboard can provide a comprehensive power system security analysis for operators.

2. **Cyber Battlefield**: This dashboard provides a platform for a Hacker who tries to compromise the system through unexpected actions [2] or bad data injection [20] and Defender who tries to save the system from the adversaries without sacrificing electricity reliability [21]. The purpose of this dashboard is to provide attacker-defender scenarios for operators' and engineers' training.

3. **GIC**: A dashboard to show the geomagnetically induced current (GIC) field intensity and the temperature data for transformers. An

alert/warning system will be enabled if the time-varying GIC electric field is available.

As an example, **Figure 3.** shows the homepage which has an updating interactive oneline diagram of the south area of the synthetic Texas grid [22, 23], real-time area data that has the total generation, the total load, the average frequency, the exported real power, and a table for bus voltage violations and a strip-chart, which is scaled to show one minute of voltage magnitude data for some key buses in the system. Each web interface has a MQTT client inside for the real-time data stream communication. By subscribing to the same topic, data received by multiple clients with different roles are synchronized, which can be seen in **Figure 2.** where notifications pop up at the same time on both operator's dashboards.

### 2.5. User Interaction

The web interface supports 21 types of interactive control action on devices in the simulated grids, which includes adjusting the voltage setpoint for the exciter, adjustable load shedding and changing reactive power support. The control action can be triggered at once, or be triggered at a specific time point, or in some applications, users are able to execute an action for an accurate time duration, for example opening a breaker for 3 cycles. Based on the roles, the operator may have limited but different operation controllability over the grid. To support the operator's decision, in addition to multiple tabular displays of the real-time measurements, the interface also has some real-time visualization to give the operator more situational awareness, as shown in **Figure 2.** where the GIC field contour shows the spatial distribution of geomagnetic impact on the grid, and the blue and green scatters show the intensity and the polarity of the GICs. A substation combobox can be triggered by clicking the substation symbols in the map to show the detailed substation data along with the substation oneline diagram to provide a finer observation and control over the grid. As shown in **Figure 4.**, operators are able to issue commands in the substation combo-box. Commands like toggling devices or changing MW setpoint can also be done in the tables where switches or editable cells are available, as shown in **Figure 7.**. Based on the role's responsibility, users may have the full capabilities to curtail loads, generations, change schedules, switch shunts and change transformer taps by clicking the buttons in the combo-box and in the tables.

Another useful feature is that after the simulation is finished, data and action records are available to download in the report page. This allows later playback by uploading the record file to the interface, which can be utilized for detailed analysis of operations. We can check whether the operator or the defense algorithm does the optimal/appropriate response regarding to the contingency.

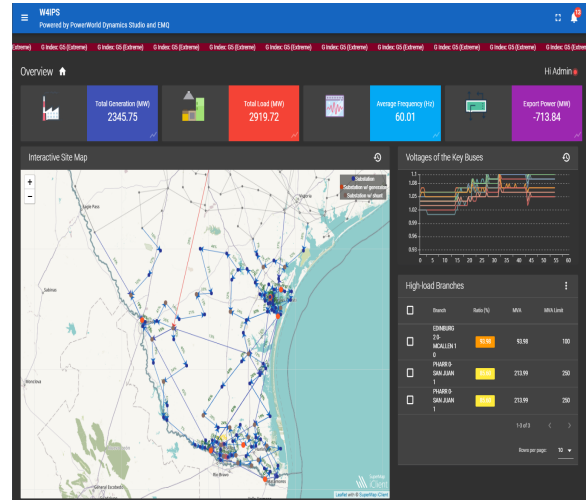

**Figure 3.** The main dashboard shows the area data on the top, a large interactive map that has a one-line diagram with real-time substation and branch data, a strip-chart on the right to show the voltage magnitudes for key buses and a table that shows the buses with voltage violations.

### 2.6. Extendibility

W4IPS can be used to not only simulate the current power grid, but also has the possibility to simulate the future power system where a huge amount of IoTs based power electronic interfaces and devices (grid-iot) running in the system. Building on top of the MQTT protocol, the real-time data pipeline ensures W4IPS has the capability to include the grid-iot devices. Though W4IPS is designed to be an interactive simulation environment, with the real-time data stream from field devices instead of DS, it can also work as an online platform for power system visualization and analysis. In addition to the transmission-level simulators that are currently used in W4IPS, it also has the capability to be used with the distribution system simulator, e.g. OpenDSS [24] and GridLAB-D [25]. The integration of the co-simulation engine HELICS [26] is also in the future development plan. Moreover, different data sources, such as real-time weather information, LMP, etc., can also be included for provide a realistic environment for power system studies and analyses.

### 3. Application of W4IPS

With the features and structure mentioned before, W4IPS has the capability for the operator training and multi-operator collaboration. In this part, we present how W4IPS can be used to help students gain experience and insights on large power system operations by exploring the inter-related operational tasks and how the W4IPS is used to train grid operators to handle GMD

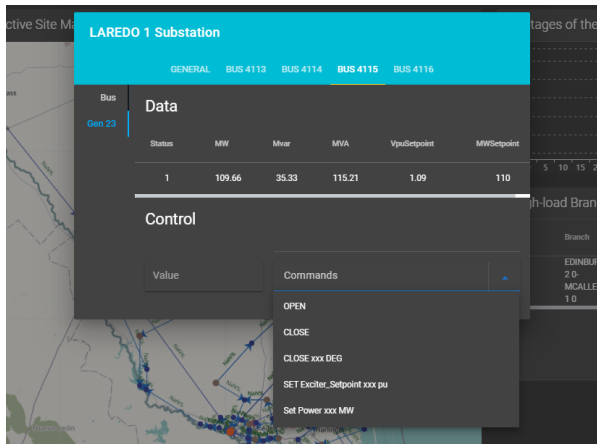

**Figure 4.** The generator data and the available control actions. Users are able to switch to other buses in the substation with the tabs on the top.

events.

### 3.1. Educational Tool for Power System Analysis

In the final lab of the senior undergraduate course "Power System Operation and Control" at Texas A&M University, students work in teams of three to control the South Texas area of the 2000-bus synthetic Texas grid. This is a brand new experience for students to collaboratively work together to operate a power grid, whereas prior labs in the course did not present the opportunity to do so [27]. The scenario runs for 10 minutes at 60 times real time, simulating peak load on a hot Texas summer day from 10:00 AM to 8:00 PM. Students interact with W4IPS to control the transmission network, generation fleet, shunt capacitor banks and if necessary, loads. Throughout the simulated day, W4IPS will report both the cost of operation in dollars and a reliability score, so each team needs to work collaboratively to operate the control area as reliably and cost-effectively as possible.

To achieve this goal, students normally will break this into three specific goals based on the roles they choose:

1. Keep the area control error (ACE) as close to zero as possible;

2. Eliminate line overflows, bus over-voltage and under-voltage violations;

3. Dynamically adjust generator outputs according to the marginal cost to decrease the total cost.

In the scenario, the AGCs are off in the South Texas area so that students have to manually adjust the power output of generators by changing the MW setpoint. Once the simulation starts, the online capacity will run out within about 200 seconds due to the rapid increase of loads. As shown in **Figure 5.**, which shows the percentage of generation for each online generator at 8:00 AM and 11:50 AM, each block stands for an online generator where the size of the block represents the capacity of the generator. If the generator is running in "low-load", the block will be green; if the real power output is about or exceeds its capacity, the block will be red. In **Figure 5.**, it is clear that the system needs more generation and capacity, so students need to close other offline generators and adjust the power output to reduce the number of overloaded lines caused by the limited generation resources. The lack of online reactive power support and the low generator terminal voltage may result in a voltage collapse or blackout, which is also a common issue that students will face.

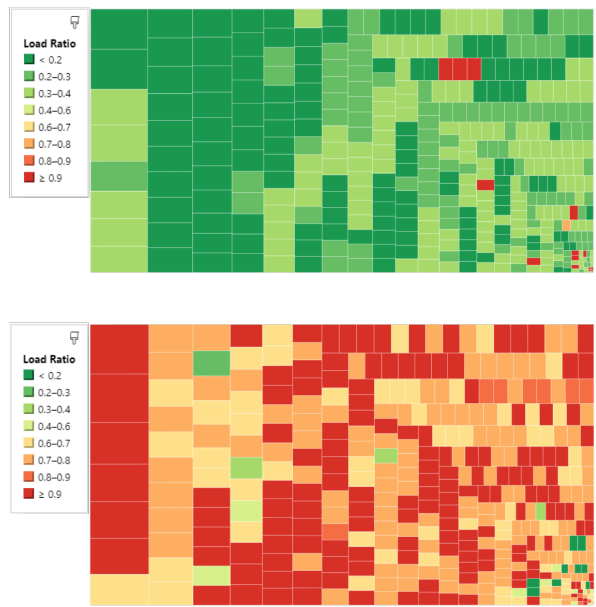

**Figure 5.** Percentage of generation for each online generator at different time: 8:00 AM (top) vs 11:50 AM (bottom). The green color stands for a low percentage of generation, while the red color represents that the generator is close or reaches its capacity.

According to our observation, students tend to have different roles when facing the simulation:

1. One student will focus on the overview page as shown in **Figure 3.** which has a dynamic chart of the load prediction for the day and the current load. The student will actively track the trend of the change of loads. In the same page there is also a geographic one-line diagram which will highlight an overflowed line.

2. One student will focus on the generator page as shown in **Figure 7.** which has a real-time pie chart for the current generation and capacity and has all operation data from generators. The student is normally busy adjusting the generator terminal voltage and the real power setpoint to match the changing load and eliminate the line overflows.

3. The last student will mainly focus on the shunt page which has a table to show the bus voltage violation. The student needs to change the shunt operation status to keep the bus voltages in a range of [0.95,1.05].

Most of the time students are able to work individually, however, when the severe situation happen, e.g. multiple line overflows and bus voltage violations occur in a short period, they will need to work together to minimize the impact. They are able to evaluate their decisions during the simulation by checking whether the reliability index improves after the actions.

Once the simulation finishes, a report with the reliability score, measurement data, total cost, violations and action records will pop up which can be used by students to evaluate their strategies or collected by an instructor for further research. Then students need to discuss among their team about what happened during this shift which helps them coordinate their tasks and get closer to the goal in the next round. The reliability score is calculated based on the numbers of the violated bus voltages and the overload branches, and **Figure 6.** shows the typical scores for the first round and fourth round. Most students will normally experience voltage collapse due to the shortage of reactive power support and the lack of online generation capacity in the first and second rounds, however they will have a better understanding for the impact of the control actions and become more sensitive to the voltage violations and power flow violations in the last few rounds.

### 3.2. Online GMD Mitigation

GMDs have the potential to severely disrupt operations of the electric grid by inducing quasi-dc GICs in the power grid, which may increase the reactive power losses and temperatures of transformers [28]. Thus, we create a scenario, which simulates an extreme geomagnetic storm event impacting the South Texas area, to give operators an immersive experience of how GMDs will affect the grid and what actions are needed to mitigate the impact.

In this scenario, an extreme geomagnetic storm alarm is received by the control center at 16:28:00, which followed by a ten-minute varying GIC electric field superimposed on the grid. As shown in the **Figure 2.**, the operator is able to closely monitor the intensity of the GIC electric field from the contour, which is generated dynamically based on the GIC electric field measurements in the substations, and keep an eye on the temperatures and the GIC neutral currents of transformers. The size of the scatter represents the intensity of the GIC neutral current at that transformer,

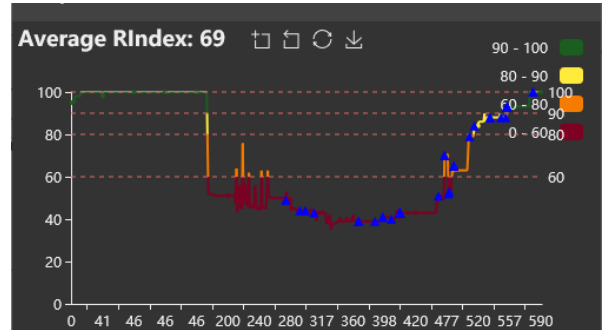

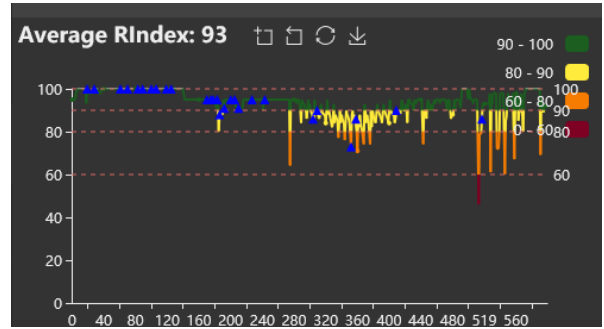

Figure 6. The typical reliability index curves for the first round (top) and the last round (bottom), where blue triangle means at that time instant there is a control action issued by the student.

while the green and blue color stand for the positiveness and the negativeness of the current.

Based on these GIC-related data and other data in the generator, load, shunt and transmission line dashboards, the operators are required to handle the voltage drop and the over-heating of transformers, which may need them to adjust the reactive power supply, change the generator output, change the grid topology by opening/closing the branches or shed loads according to the severeness of the impact.

### 4. Future Work with W4IPS

For power system security studies, with its cyber-physical nature, it is necessary to include the set of control, communication, and physical system components. The pure power system simulation can provide SA from power system perspective, but it lacks the information from the cyber perspective and the device level. Several testbeds has been proposed for the analysis of cyber-physical security in power systems with the integration of power system simulation, communication network emulation, physical device and real-time data in [29–32]. However, an important element of power system security is the human factor, which is hard to quantify and test in CPS.

**Figure 8.** shows the framework of integrating W4IPS with a Hardware-In-the-Loop (HIL)

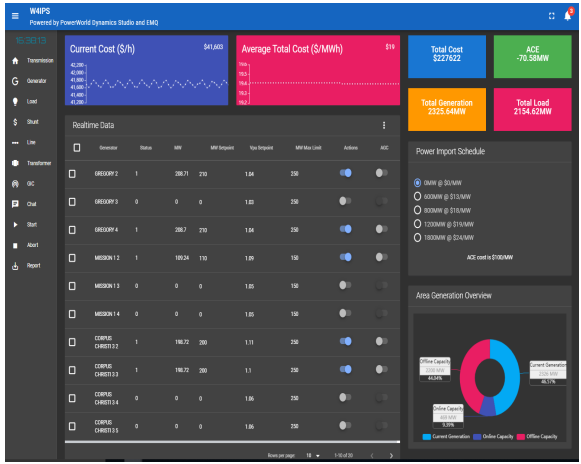

Figure 7. Generator operator's display in W4IPS. On the top there are two charts showing the current cost and the average cost. A table shows the real power and the terminal voltages, with functions to change the real power setpoint and the voltage setpoint. On the right there is area generation data, a power import option and a pie chart that shows the online capacity, the offline capacity and the current generation.

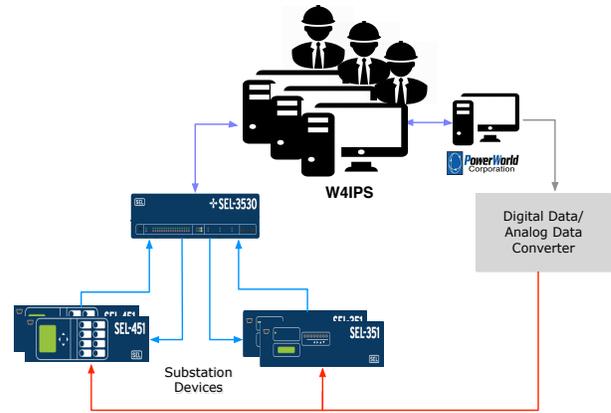

Figure 8. Future application with W4IPS in the HHIL testbed

testbed to introduce human factors into the power system cyber security analysis, making it a Human-Hardware-In-the-Loop (HHIL) testbed. The interactive action between operators and simulation allows the test and verification of operators SA, judgement and experience to different contingencies. The hardware allows a specific focus on the operation and function of the target device. With the analyses from both cyber and physical perspectives as in [18, 19], it will inform operators how important that device is and how to secure the device. Based on the real-time power system data from W4IPS, power system security can be analyzed from the compromised devices, malicious operations in power systems, operators' SA and reactions to contingencies, etc. With a complete view of power grid, targeted device, and operators response to contingencies, the HHIL testbed can be used for power system security analysis, training and test.

Whenever there is a cyber induced physical hazard in a device or adversaries in a power system simulation, it can be observed in W4IPS. With the interactive control functionality, the devices can be directly controlled by different operators. This feature makes the W4IPS-integrated HHIL testbed an ideal platform to test the defense and attacker algorithms for power system cyber-physical security studies, as shown in **Figure 9.**. On one side, defenders need to ensure the stability and reliability of power systems under attacks from adversaries by shedding loads, adjusting the output of generators, or islanding specific areas to avoid cascading outages and ensuring overall power supply. On the other side, attackers may try to compromise the devices, inject bad data to the simulation, or mis-operate critical elements to sabotage the whole system. With the multi-user feature and web interface, users can join the testbed remotely and play different roles simultaneously to have a real-world power system attacker-defender competition.

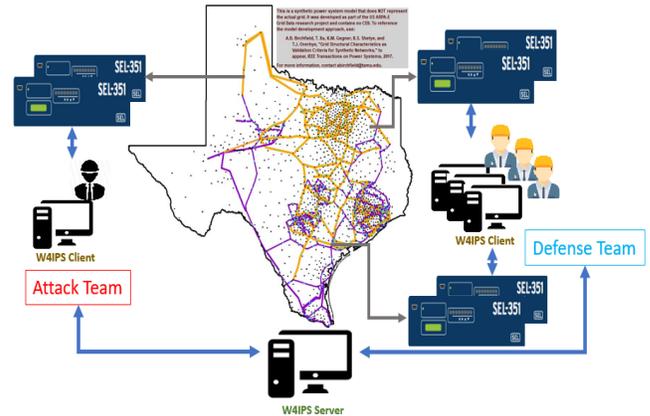

Figure 9. The proposed environment is capable of running the cyber competition between defenders and attackers.

Like the lab mentioned previously, operators SA can also be expressed through the W4IPS through their judgments of a power system status. Different operators may choose different contingency response strategies to a contingency at different timing. The following power system status of their operation can inform them whether they make the optimal decision or what information they need the most to make the decision at the early stage. Thus, W4IPS can be utilized as a training and evaluating tool for operators to test whether they are aware of potential threats in their power systems and they can make the right decision to save the system in a real-time manner. Moreover, the synthetic power system topology information, cyber-physical architecture information, and network traffic data can also be transmitted to the W4IPS with its extendibility, which makes the W4IPS a centralized location for all power system information. Such information can be utilized for various security analyses. The combination of W4IPS and other security analysis platforms, like CyPSA [18], can provide a comprehensive understanding of power systems security analysis with the real-time visualization and cyber-physical security assessment.

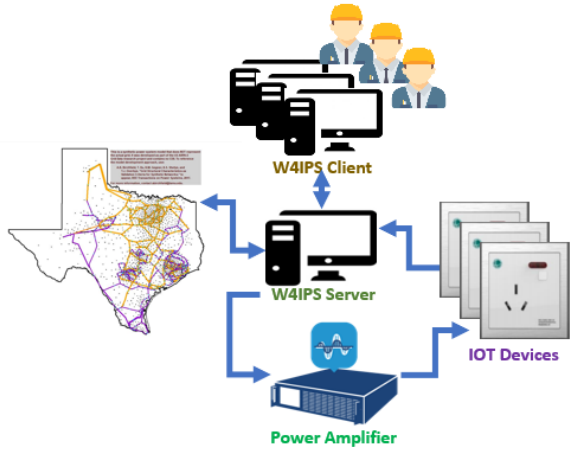

**Figure 10. The proposed framework integrates the IOT devices into the HHIL testbed.**

Other applications with hardware involved can also be done based on W4IPS. For example, W4IPS can be easily used for simulating the future grids with the integration of IoT devices in distribution and user level, because W4IPS and most IoT devices share the same MQTT-based communication protocol and structure. The distribution system can be modeled in a different platform with more details, like feeder information, household information, etc [24, 25]. As shown in **Figure 10.**, the source-grid-load interactivity can be simulated in the extended HHIL environment. Ultimately, the impact of this grid-controllable devices [33] and the vulnerabilities of the underlying cyber network can be further analyzed. With large amounts of data and large-scale systems, W4IPS can also be utilized to provide a platform for developing multi-agent using deep reinforcement learning to operate the grid with human operators through different data sets for corresponding contingency, power grid status and operators' response.

## 5. Source Code of W4IPS

The W4IPS is an open source project with the MIT license [34], thus it can be utilized to do the similar tasks as mentioned in the use case, or modified for other applications. For the education and GMD use cases, users do not need to install any dependencies and are able to experience the system by browsing our website directly. To modify the system to have other functionailities or use it with other simulation engines, the user could clone the repository, install the dependencies and then modify the code based on the needs. The simulation engine in this paper is not included in the license, however users are able to request a demo version through their website [35].

## 6. Conclusion

This paper has presented a web-based interactive power system dynamics simulation environment for power system security analysis, which is capable of visualizing the real-time data, performing dynamic simulations, and introducing human factors with the multi-user interactivity. The web-based interface allows people from different areas to collaboratively operate on the same grid to control the devices in the system.

The applications of W4IPS shows its ability to provide a comprehensive understanding of power systems and how operators' decision can affect the system. The real-time visualization provides a better understanding of power systems. The interactive control functionality makes W4IPS a platform suitable for the response operation under different contingency scenarios. For the future work, the W4IPS can be connected with different testbeds for further cyber-physical security studies for power systems. With other cyber-physical security analysis toolsets, the W4IPS can provide a comprehensive security analysis and validation. The W4IPS can also be extended to use at IoT-embedded grid studies and power system AI development.

## Acknowledgment

The authors would like to thank the US Department of Energy Cybersecurity for Energy Delivery Systems program under award DE-OE0000895 and National Science Foundation under Grant 1520864.